\documentclass[%
twocolumn,
superscriptaddress,
 amsmath,amssymb,
 aps,
]{revtex4-2}

\usepackage[english]{babel} 
\usepackage[T1]{fontenc}
\usepackage[ansinew]{inputenc}

\usepackage{amsmath}
\usepackage{amsthm}
\usepackage{amsfonts}
\usepackage{bbold}
\usepackage{xcolor}
\usepackage{soul}
\usepackage{cancel}
\usepackage{color}

\usepackage{graphicx}
\usepackage{dcolumn}
\usepackage{bm}

\newcommand{\M}{\mathbf{M}}
\newcommand{\m}{\mathbf{m}}
\renewcommand{\r}{\mathbf{r}}
\renewcommand{\H}{\mathbf{H}}

\begin{document}

\preprint{v000 - \today}

\title{Blind in-memory arithmetic with confined skyrmions: doubling and successor functions on a skyrmion byte}

\author{Joan Revert\'e}
\affiliation{Departament de F\'isica, Universitat Aut\`onoma de Barcelona, 08193 Bellaterra, Barcelona, Spain}

\author{Nuria Del-Valle}
\affiliation{Departament de F\'isica, Universitat Aut\`onoma de Barcelona, 08193 Bellaterra, Barcelona, Spain}

\author{Jaume Cunill-Subiranas}
\affiliation{Departament de F\'isica, Universitat Aut\`onoma de Barcelona, 08193 Bellaterra, Barcelona, Spain}

\author{Natanael Bort-Soldevila}
\affiliation{Departament de F\'isica, Universitat Aut\`onoma de Barcelona, 08193 Bellaterra, Barcelona, Spain}


\author{Carles Navau}
\email{carles.navau@uab.cat}
\thanks{Serra H\'unter Fellow}
\affiliation{Departament de F\'isica, Universitat Aut\`onoma de Barcelona, 08193 Bellaterra, Barcelona, Spain}


\begin{abstract}
We present a skyrmionic-based computation platform, able to perform fully in-memory computing by collectively manipulating the skyrmions that serve as bits of the binary memory. The computation protocols are implemented through driving currents and adequately located anisotropy defects that modify the dynamics of the skyrmions along a magnetic track. Simulations demonstrating the feasibility of such an in-memory computing system are shown for several basic and simple mathematical operations with natural numbers, such as doubling and adding one (the successor function). We also discuss how further optimization of the geometry, barrier profiles, current amplitudes, and timing parameters could improve both the robustness and the overall practical feasibility of the system.
\end{abstract}


\maketitle

\section{Introduction}
\label{sec:intro}

The accelerated growth of processing digital data is pushing basic research and technology development to new paradigms in computation. One of the principal problems is the so-called von Neumann bottleneck, that stems from the physical separation between the memory and the processing units. The amount of information transiting to and from these two units consumes a large amount of the energy and time of the computation \cite{Mutlu2019-Micro}.

In-memory computing is the name given to the computation paradigm in which the computation of data is done in-situ, within the memory bits themselves \cite{Ielmini2018-NatElectron,Pedretti2021-Electronics,Sebastian2020-NatNano}, thus avoiding most of the energy costs of reading back and processing the memory content and the associated latency. Various levels of integration of memory/computation are being proposed, the most radical being the total integration in which the computation is done without the need of reading the memory at any step of the computation \cite{Mannocci2023-APLMachLearn,Sebastian2020-NatNano}. 

One of the recent candidates to be used as units of magnetic memories are magnetic skyrmions. Indeed, magnetic skyrmions are swirling magnetic structures whose potential for applications comes mainly from three important properties \cite{Everschor-Sitte2018-JAP,Fert2017-NatRevMat,Schutte2014-PRB,Nagaosa2013}: they are small (skyrmions of sizes of about 15 nm have been observed \cite{Caretta2018-NatNano}); they can be moved by current densities of the order of 10$^6$ A/m$^2$ for bulk B20 compounds \cite{Jonietz2010-Science,Schulz2012-NatPhys} or $10^{10}-10^{11} \,\mathrm{~A/m^2}$ for ultrathin films with thickness $\simeq 10\,\mathrm{~nm}$ \cite{Woo2016-NatMat,Litzius2017-NatPhys}; and they are topologically protected, which means that they are stable under small perturbations, even at room temperature. With these properties, one can consider the presence/absence of a skyrmion as a 1/0 bit of information and think of using them as information carriers. A number of proposals using skyrmionic devices as carriers of information have been presented: (i) skyrmionic-based racetrack memories, where skyrmions replace the magnetic domains in a non-volatile information storage device in the form of tracks \cite{Fert2013-NatNano,Parkin2008-Science}, (ii) skyrmion-based logic computing gates, where a couple of skyrmions enter a specifically designed system such that the interactions between them, together with the external driving currents and the adequate border forces, allow one to design systems behaving as logic gates \cite{Zhang2015SciRep,He2017-IEEEMagLet,He2024-PRAppl,Mankalale2019-IEEETransElecDev,Luo2018-NanoLett,Chauwin2019-PhysRevAppl};  or (iii) skyrmion-based transistor-like functional devices in which the transport of skyrmions is controlled by a gate voltage, mimicking a transistor function \cite{Zhang2015SciRep}. 

Skyrmion-based devices have been proposed for unconventional computing schemes: (i) Neuromorphic computing devices, inspired by the biological brain, where skyrmions were used to design an artificial synapse \cite{Huang2017-Nanotechnol,Song2020,Grollier2020}, (ii) reservoir computing devices, where confined skyrmions would be the required nonlinear-in-time elements of the reservoir having spiking behaviour \cite{Prychynenko2018,Lee2023};
(iii) random bit generators, where the stochastic behavior of skyrmions in presence of thermal fluctuations allows the generation of random bits to be achieved \cite{Medlej2020,Navau2021-APLMat}, or (iv) stochastic computing, where skyrmionic chains can be reshuffled \cite{Zazvorka2019-NatNano,Pinna2018}.

Here we focus on the in-memory computing potential of magnetic skyrmionic platforms. We shall describe a system in which confined skyrmions are considered bits of information and, at the same time, their controlled movement along a magnetic track ---where several defects are adequately introduced--- is interpreted as an arithmetic computation function. The main idea is to provide protocols for fully blind computations: without knowing where the bits are (the position of the skyrmions), we shall find that their change in position after applying the protocol will correspond to a given mathematical operation [$n \longrightarrow f(n)$]. The physical implementation of the computation protocols will be introduced through driving currents and adequately located anisotropy defects that modify the dynamics of the skyrmions along the track \cite{Gonzalez-Gomez2022-PRAppl,Navau2018-JMMM,Castell-Queralt2019-Nanoscale}. 
In other words, given a set of bits ---presence/absence of skyrmions--- along a track corresponding to a numerical value $n$, we aim to perform mathematical operations, $f(n)$, without the need to know the value of each bit. That is, to perform a real in-memory computing without the need of reading/writing the information. Several works have been devoted to the use of skyrmions in logic memories \cite{Zhang2020-PRAppl,Zhang2015SciRep,Luo2018-NanoLett,Yan2021-PRAppl}, although far fewer are devoted to direct blind computation platforms using skyrmions. Worth noting are the works of Sisodia et al. \cite{Sisodia2022PRAppl} and Gnoli et al. \cite{Gnoli2021-Electronics} where a skyrmionic full-adder (that can be reconfigured to perform different logic operations) and a max/min seeker were designed, respectively. In these cases, the operation is done at the bit level, just adding or comparing the values of two bits to generate an output. In our case, the key point is the blind global operation. What we propose is a genuine in-memory computing: given a set of bits in a portion of magnetic memory, they are modified in such a way that the final memory bits correspond to the result of the desired operation.

Naturally, there are infinitely many possible operations. We shall start with the simple example of the multiplication by $2^m$, $P(n)=2^m n$. Next, we focus on the successor function $S(n)=n+1$, which, together with the constant $0$, generates by composition and primitive recursion all primitive recursive functions, including addition, multiplication, and exponentiation \cite{Boolos_Burgess_Jeffrey_2007}. Many further operations could be defined from combinations of basic functions. Our aim here, however, is to introduce only the most fundamental ones, emphasizing their definability and theoretical possibility, rather than considerations of computational efficiency.

The paper is structured as follows. In section \ref{sec:platform} we introduce the skyrmionic system used. In particular we introduce the bit-cell and the byte-track as the basic geometric elements in which the skyrmions will be hosted. In section \ref{sec:manipulation} we present basic manipulation actions we can use to reorder the skyrmions along the byte-track. In particular we introduce the defect barriers and the driving torques used to manipulate skyrmions. In section \ref{sec:operations} we describe the different protocols performing in-memory computing for especially relevant operations, such as multiplication by 2 and +1 addition (successor). In section \ref{sec:discussion} we discuss the validity and performance of the described system and present some conclusions.

\section{The skyrmionic in-memory computing platform}
\label{sec:platform}
\begin{figure}
    \centering
    \includegraphics[width=1\linewidth]{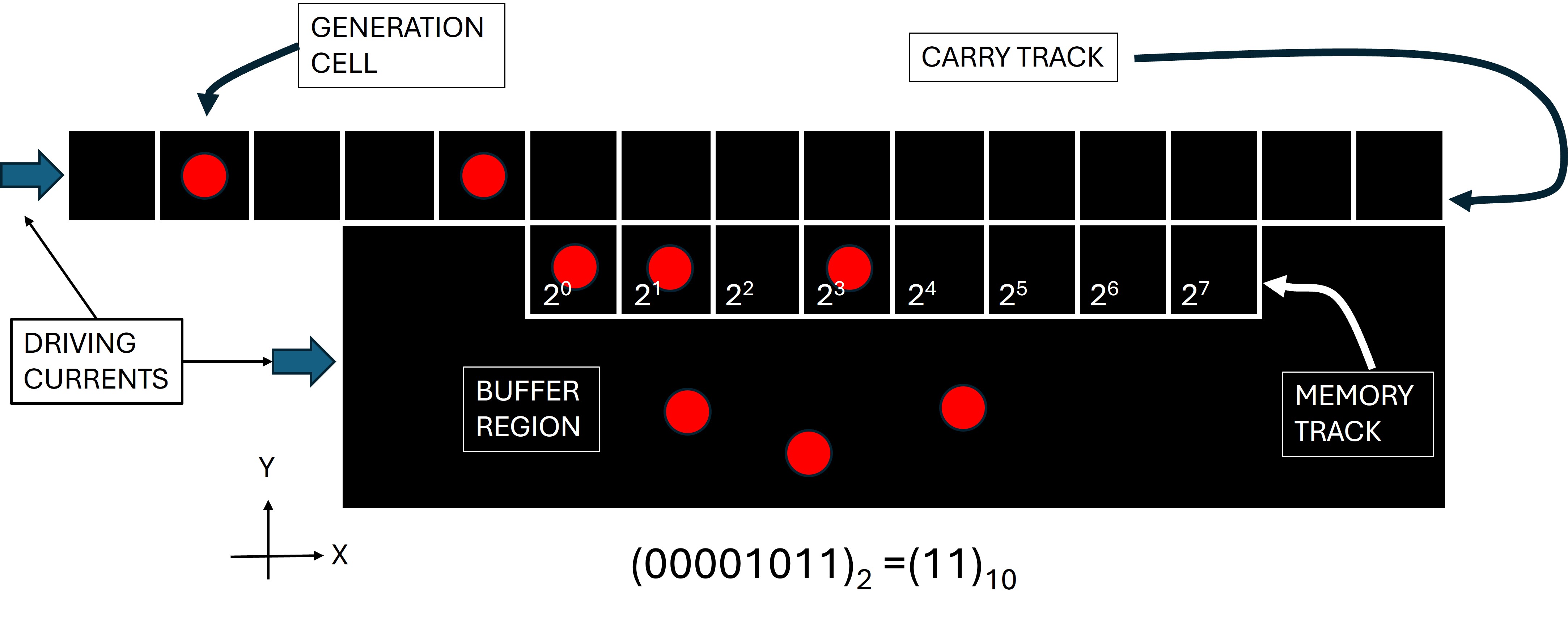}
    \caption{Sketch of the considered skyrmionic in-memory computing platform. The magnetic memory is stored in the memory track. The bit-cells are surrounded by defect lines representing a local increase of the anisotropy constant with respect to the anisotropy of the rest of the sample. Red circles represent skyrmions, a "1" bit in the memory. The number in each bit-cell of the memory-track indicates the value of the bit-cell in ordinary decimal basis (note that the least significant bit is on the leftmost cell of the memory track). The carry region will be used for controlling the skyrmionic flux. The buffer region can contain skyrmions that would be eventually destroyed. The bold arrows represent the current fed in the substrate to produce the spin-polarized current that drives the skyrmions. In the example, the memory track codes the binary number 00001011, corresponding to 11 in decimal. The carry track contains two skyrmions while in the buffer track there are three skyrmions at arbitrary positions.}
    \label{fig:sketch}
\end{figure}

We consider an ultrathin (thickness $d$, much smaller than other dimensions of the sample) ferromagnetic (FM) sample 
located on the plane $z=0$ on top of a heavy metal substrate. 
The FM is divided into three regions (see Fig.~\ref{fig:sketch}): (i) a carry-track formed by a longitudinal set of bit-cells of size 50$\times$50 nm$^2$ each, separated by anisotropy defects. In our simulations we will consider 15 of such cells and, when needed, the skyrmions will be generated in one of those bit-cells (the one marked as generation cell), this carry-track forms the top-region; (ii) below the carry-track, a "memory-track" formed by eight 50$\times$50 nm$^2$ cells separated from the carry-track and between themselves also by anisotropy defects. Each of these cells represents a bit of magnetic information, where the presence or absence of a skyrmion indicates 1 or 0, respectively; (iii) finally, below and surrounding the memory-track, there is a buffer region, where remaining skyrmions can be temporarily stored or eventually destroyed. 

The heavy metal substrate is also considered as formed by separated tracks allowing currents to be injected in the different regions independently.

For simplicity, we will constrain ourselves to a memory-track with 8 cells corresponding to a byte of information. Thus, we are considering non-negative integers smaller than 2$^8$ = 256. The protocols can be extended directly to larger numbers by adding bit-cells, and, with an appropriate encoding (e.g., two's complement or fixed point), they can also handle negative and non-integer numbers \cite{parhami2009computer}.

The micromagnetic model is used to evaluate the magnetization distribution $\M(\r)=M_s \, \m(\r)$ all over the considered sample, solving the Landau-Lifshitz-Gilbert equation
\begin{equation}
    \frac{{\mathrm{d} \M}}{\mathrm{d}t} = -\gamma \M \times \H_{\rm eff} + \frac{\alpha }{M_s} \M \times  \frac{{\mathrm{d} \M}}{\mathrm{d}t}+ \mathbf{T},
\end{equation}
where $\alpha$ is the Gilbert damping parameter, $\gamma$ the gyromagnetic ratio ($\gamma=2.21\times 10^5$~m~A$^{-1}$s$^{-1}$), $M_s$ is the saturation magnetization, and $\H_{\rm eff}$ the effective field. $\H_{\rm eff}$ contains the exchange, effective out-of-plane uniaxial anisotropy, interfacial  Dzyaloshinskii-Moriya (iDM) interactions. No applied field is considered. The demagnetizing fields are included in the effective anisotropy term as the sample is an ultrathin film \cite{Gonzalez-Gomez2022-PRAppl}. $\mathbf{T}$ is the torque exerted over the magnetization due to eventual polarized currents. These currents will be used to create skyrmions and to drive their movement. 

For the numerics, we have considered typical values of a CoPt system \cite{Sampaio2013-NatNano,Gobel2019-PRB,Gonzalez-Gomez2022-PRAppl}: $M_s=580\,\mathrm{~kA/m}$, exchange constant $A=15 \, \mathrm{~pJ/m}$, iDM constant $D=3.0 \,\mathrm{~mJ/m}^2$, effective uniaxial anisotropy $K_0=4.25 \times 10^{5}~\mathrm{J/m}^3$, and $\alpha=0.1$. With these values, the exchange length is $l_{\rm ex}=\sqrt{(2A)/(\mu_0M_s^2)}=8.42\,\mathrm{~nm}$, and we divide the sample in cells of side $\Delta=2.5\, \mathrm{~nm} \simeq l_{\rm ex}/3$, and thickness $d=2.0~\mathrm{nm}$.
The defects, always located at the borders of the cells, are simulated as local variations of the anisotropy constant \cite{Castell-Queralt2019-Nanoscale,Del-Valle2022-APLMat,Everschor-Sitte2017-NewJPhys}. In particular, as baseline, we consider that the left-right borders (vertical lines in Fig.\ref{fig:sketch}) have a value $K=10K_0$ and the  upper-bottom borders (horizontal lines in Fig.\ref{fig:sketch}) have a value $K=3K_0$. These values will be locally and temporarily changed to control the movement of the skyrmions, as explained in next sections.

To evaluate the presence or absence of a skyrmion, independently of their deformation inside a cell, we evaluate the cell-skyrmion-number $N_{{\rm sk},c}$, as the integration over a given cell $c$ of the density of skyrmionic number. That is, we evaluate 
\begin{equation}
    N_{{\rm sk},c}=\frac{1}{4\pi \, M_s^3} \int_{S_{c}} \M \cdot \left(\frac{\partial \M}{\partial x}\times \frac{\partial \M}{\partial y}\right) \,\, {\rm d}S,
\end{equation}
where $c$ indicates the evaluated cell and $S_c$ its surface. If  $N_{{\rm sk},c}$ in a given cell is close to unity, a skyrmion is present in that cell, while if $N_{{\rm sk},c}$ is close to zero there is no skyrmion in that cell. As an operational criterion, we have considered that a skyrmion is in a given cell $c$ when its $N_{{\rm sk},c}>0.8$. In all cases, we checked that these values really correspond to a (possibly deformed) skyrmion. When $N_{{\rm sk},c}<0.2$ the distortion from the uniformity ($N_{{\rm sk},c}=0$) is due to boundary effects that slightly modify the uniform distribution, but not to the presence of a skyrmion. The total skyrmionic number over all the cells is $N_{\rm sk}=\sum_c N_{{\rm sk},c}$.

\section{Primitive manipulation tools}
\label{sec:manipulation}
\subsection{Creation of a skyrmion}
The generation of skyrmions has been previously studied in the literature. For example, the presence of notches in the structure \cite{Koshibae2015,Iwasaki2013-NatNano} or approaching non-uniform magnetic fields \cite{Del-Valle2015} trigger the generation of skyrmions. However, one of the preferred approaches is to use a spin-valve device through which an injected current is spin-polarized and then, after flowing through the magnetic track, the magnetization is distorted and the topological barrier can be overcome, generating a skyrmion if the conditions on the ferromagnetic track favor it \cite{Kang2017,Zhang2015SciRep,Sampaio2013-NatNano}. In our in-memory track, we use a similar approach, although the presence of the close borders reduces the range of parameters in which a skyrmion can be generated. We consider the following generation scheme. We inject an out-of-plane current $J_{T}$ in a $20\mathrm{~nm}$ diameter circular region, centered on the generation cell during a given time interval $\Delta t_T$. The effect of this current is evaluated by a torque term given by \cite{Sampaio2013-NatNano,Gonzalez-Gomez2019-PRB,Wang2019-JMMM}
\begin{equation}
{\bf T}_{T}=\frac{\mu_B P J_{T}}{e \, d\, M_s^2} \M\times(\m_p\times\M).
\end{equation}
$\m_p$ is the direction of the magnetization of the fixed layer (set at $\bf{\hat z}$), $P$ is the polarization factor, $\mu_B$ is the Bohr magneton, and $e$ is the electron charge ($e>0$). After this, the injection is switched-off and the system is left for a long time to ensure that the skyrmion, if created, remains stable. 

\begin{figure}
    \centering
    \includegraphics[width=1.0\linewidth]{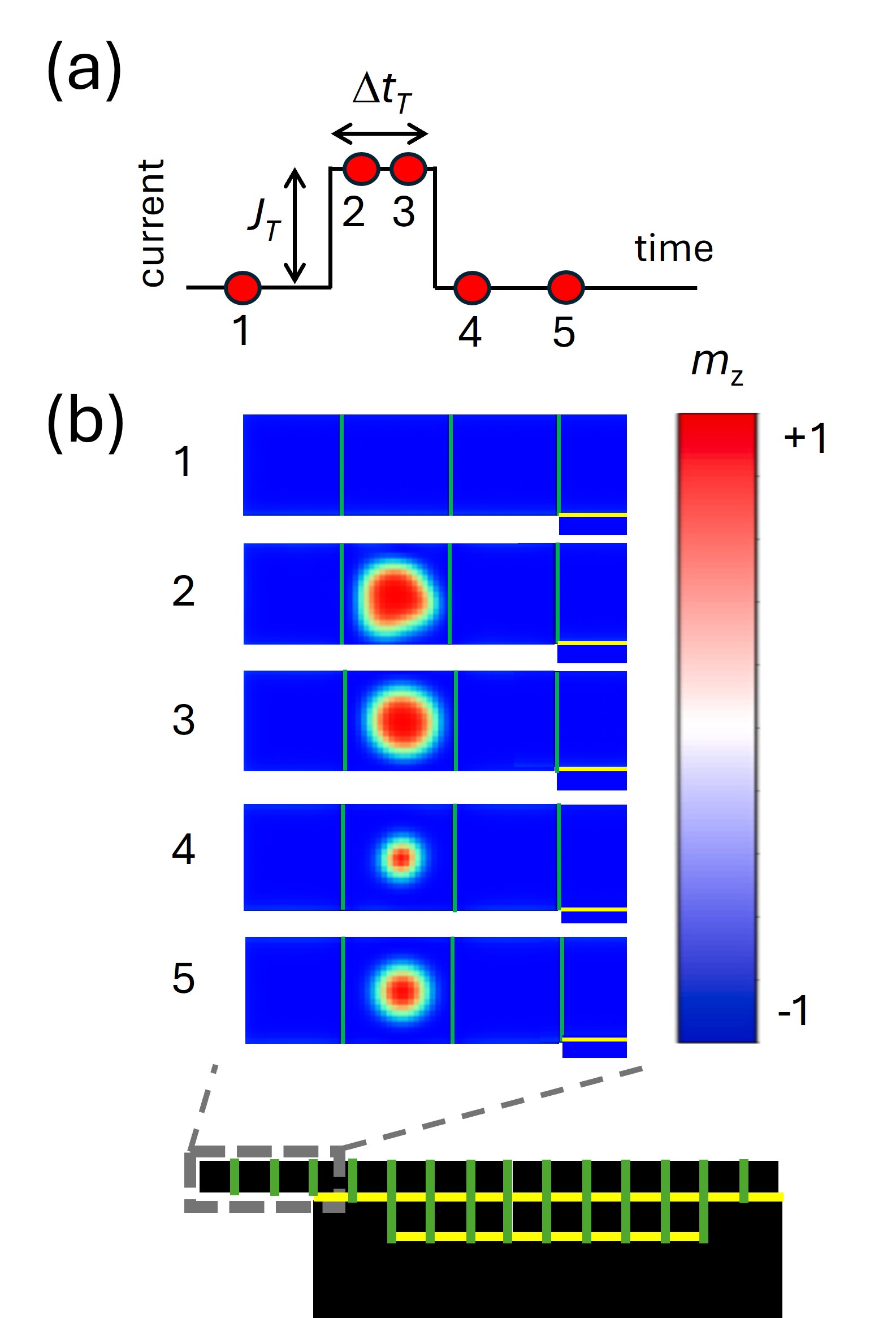}
    \caption{(a) Schematics of the protocol used for the creation of the skyrmion, the solid line sketches the value of the current as a function of time, and the red dots indicate the instant selected for the snapshots of (b). (b) Snapshots of the $z$ component of the normalized magnetization distribution corresponding to different instants of skyrmion formation. Only the indicated zoomed part of the whole sample is shown.}
    \label{fig:generation}
\end{figure}

We show in Fig. \ref{fig:generation} snapshots of the generation of a skyrmion in the generation cell using ${\bf m}_p = \hat{\bf z}$, $P=0.4$, $J_T = 2.85\times 10^{13} \, \mathrm{~A/m^2}$, and $\Delta t_T =  0.1\, \mathrm{~ns}$. The study of the phase diagram for a stable creation of a skyrmion in this particular confined geometry is beyond the scope of the present work and will be presented elsewhere. 

\subsection{Controlled skyrmion movement}
\begin{figure*}
    \centering
    \includegraphics[width=1\linewidth]{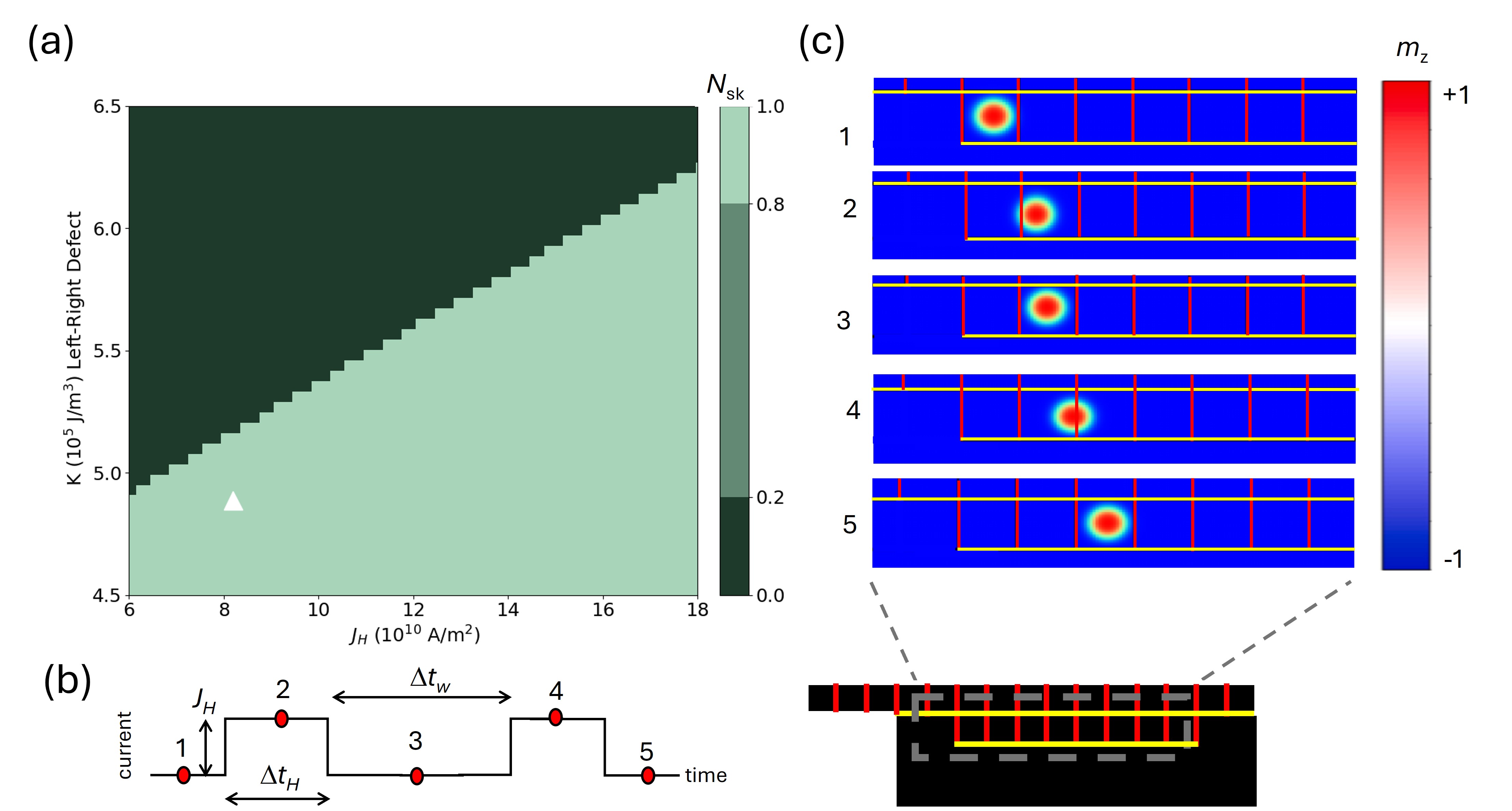}
    \caption{(a) Phase diagram for the movement of skyrmions along the memory track for different driving current and defect strength of the left-right borders of the cell. (b) Schematic time sequence for the protocol followed to move one skyrmion. (c) Snapshots of the normalized $z$-component of the magnetization distribution corresponding to the times indicated in red circles in (b); the values used for the snapshots are indicated with a triangle in (a).}
    \label{fig:1skyrmion}
\end{figure*}

Injecting a current $\mathbf{J}_H=J_H \hat{\mathbf{x}}$ in the heavy metal substrate produces a spin-polarized current through the ferromagnet that produces a torque \cite{Liu2012-PRL,Gonzalez-Gomez2019-PRB}
\begin{equation}
    \mathbf{T}_H = -\frac{\mu_B \theta_H J_H}{e M_s^2 d} \M \times (\M \times \hat{{\mathbf{y}}}),
\end{equation}
where $\theta_H$ is the spin Hall angle factor (set at $\theta_H=0.3$). It is well known that this torque produces a movement of the skyrmions along the ferromagnetic system \cite{Fert2013-NatNano,Tomasello2014-SciRep,Iwasaki2013-NatNano,Sampaio2013-NatNano}. Due to the skyrmionic Hall effect the skyrmion does not naturally move in the direction of the current. However, below a certain current threshold, the borders can guide the skyrmions along the desired direction \cite{Navau2016-PRB,Castell-Queralt2019-Nanoscale}. This property is used to move and guide the skyrmions through the different cells. The cell borders act as barriers that the skyrmions could surpass or not depending on the value of the current $J_H$ and on the strength of the anisotropy that defines the cell borders. We show in Fig. \ref{fig:1skyrmion}(a) which $J_H$ and anisotropy $K$ values of the left-right borders of the cell allow the movement of the skyrmion along the track (the upper-bottom borders are left at $K=3K_0$). For this simulation, we have considered (i) that a skyrmion is already in a given bit-cell, (ii) that the upper-bottom anisotropy is enough to avoid that the skyrmion passes through these borders, (iii) that a current $J_H$ is applied during a time interval $\Delta t_H$, after which a time $\Delta t_w$ is left to allow stabilization, before injecting again the current. In Fig. \ref{fig:1skyrmion}(c) we show several snapshots corresponding to different time moments of the protocol.

The movement of one skyrmion from cell to cell can be, thus, controlled with a quite wide range of parameters. In the following, as standard values for driving, we shall consider $J_H=8.19\times 10^{10}$ A/m$^2$. For the plots in Fig. \ref{fig:1skyrmion} $\Delta t_H = 1 \,\mathrm{ns}$, $\Delta t_w=2\,\mathrm{ns}$, and $K(\mathrm{left-right)=1.15K_0}$ have been used.

\subsection{Barrier}

The next manipulation tool to use is the local and temporal modification of border anisotropy barriers. Physically, this could be achieved incorporating local variations of anisotropy induced by local voltage-controlled magnetic anisotropy \cite{Olleros-Rodriguez2025-AdvPhysRes,Rana2024-JAP,Kang2016-SciRep}. Here we consider that we can change the value of the local anisotropy in some of the cell-divisions by applying or releasing a local voltage. In this case, the different barriers can be manipulated so that one can implement a gate-like mechanism that can be opened or closed to regulate skyrmion flow. 

To show how it works we consider that a skyrmion is on the carry track and can or cannot enter the memory track. 
To quantify this possibility the skyrmionic number in a given cell in the memory track, say cell $i$, is evaluated. Consider that a skyrmion is on the carry track, at position $i-1$, ready to enter or not to the $i$-th memory cell. We locally and temporarily reduce the left border anisotropy of the $i$-th carry cell from $10K_0$ to $1.15K_0$, while the anisotropies of the upper and bottom borders of the $i$-th cell of the memory track are reduced from $3K_0$ to $1.45K_0$ and $1.25K_0$, respectively. The values of the new anisotropies and the interaction between the present skyrmions will determine the overall behavior.

\begin{figure*}
    \centering
    \includegraphics[width=1\linewidth]{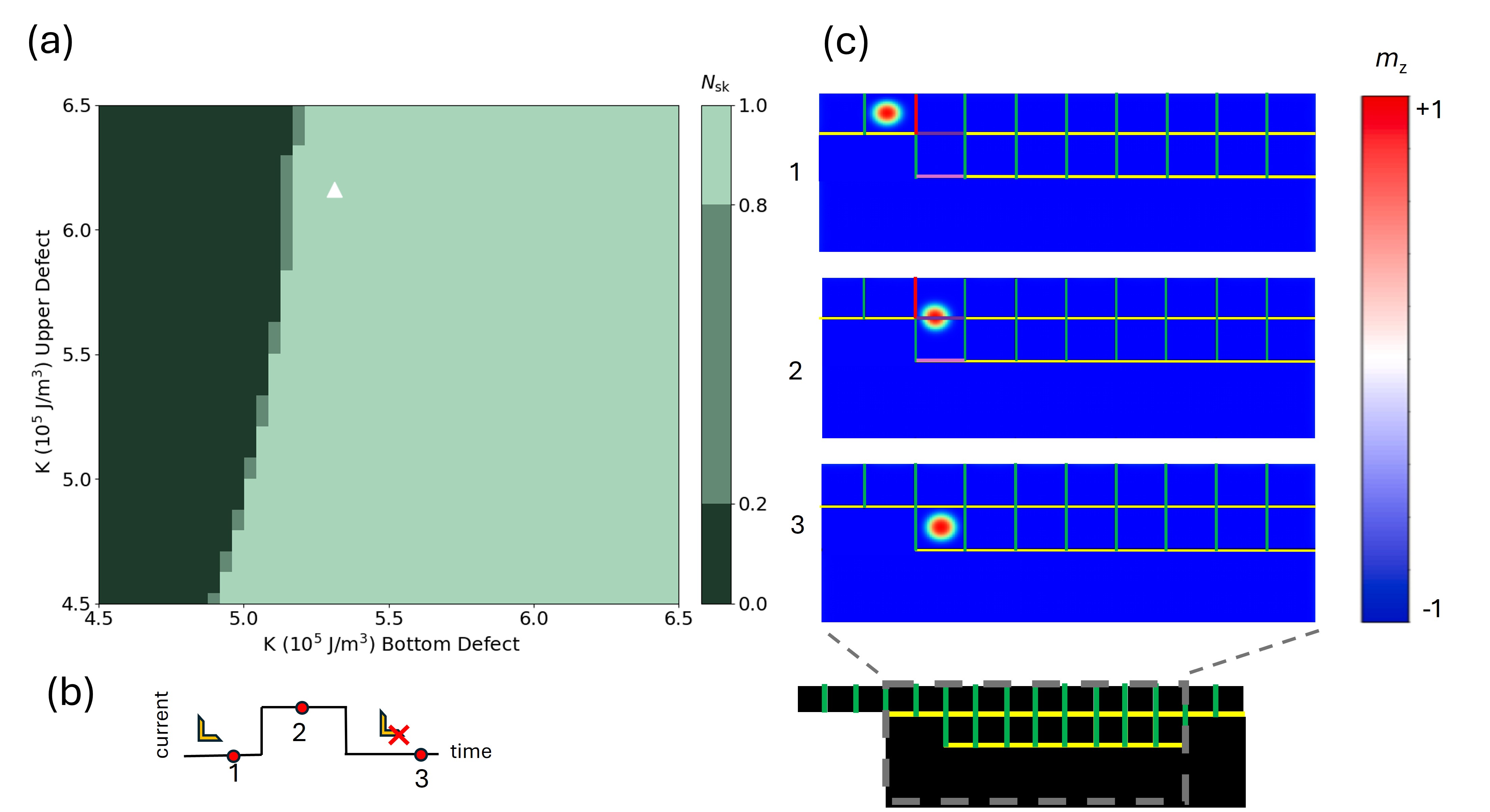}
    \caption{(a) Phase diagram for the skyrmionic number in a memory cell as a function of the values of the anisotropy of the upper and bottom borders of the $0$-th memory cell; the left border of the $0$-th carry cell is also modified at $1.15K_0$; initially, there are no skyrmions in the memory cells. (b) Sketch of the temporal protocol applied; the symbology is the same as the sketch in Fig.\ref{fig:1skyrmion}; the symbol {\huge{${\bf \llcorner}$} }indicates activation ---deactivation, when crossed out--- of the border anisotropy reduction from the baseline values. (c) Snapshots of the $z$-component of the normalized magnetization at the instants indicated as red points in (b) using the values for anisotropies indicated with a triangle in (a).}
    \label{fig:barrier1}
\end{figure*}

\begin{figure*}
    \centering
    \includegraphics[width=1\linewidth]{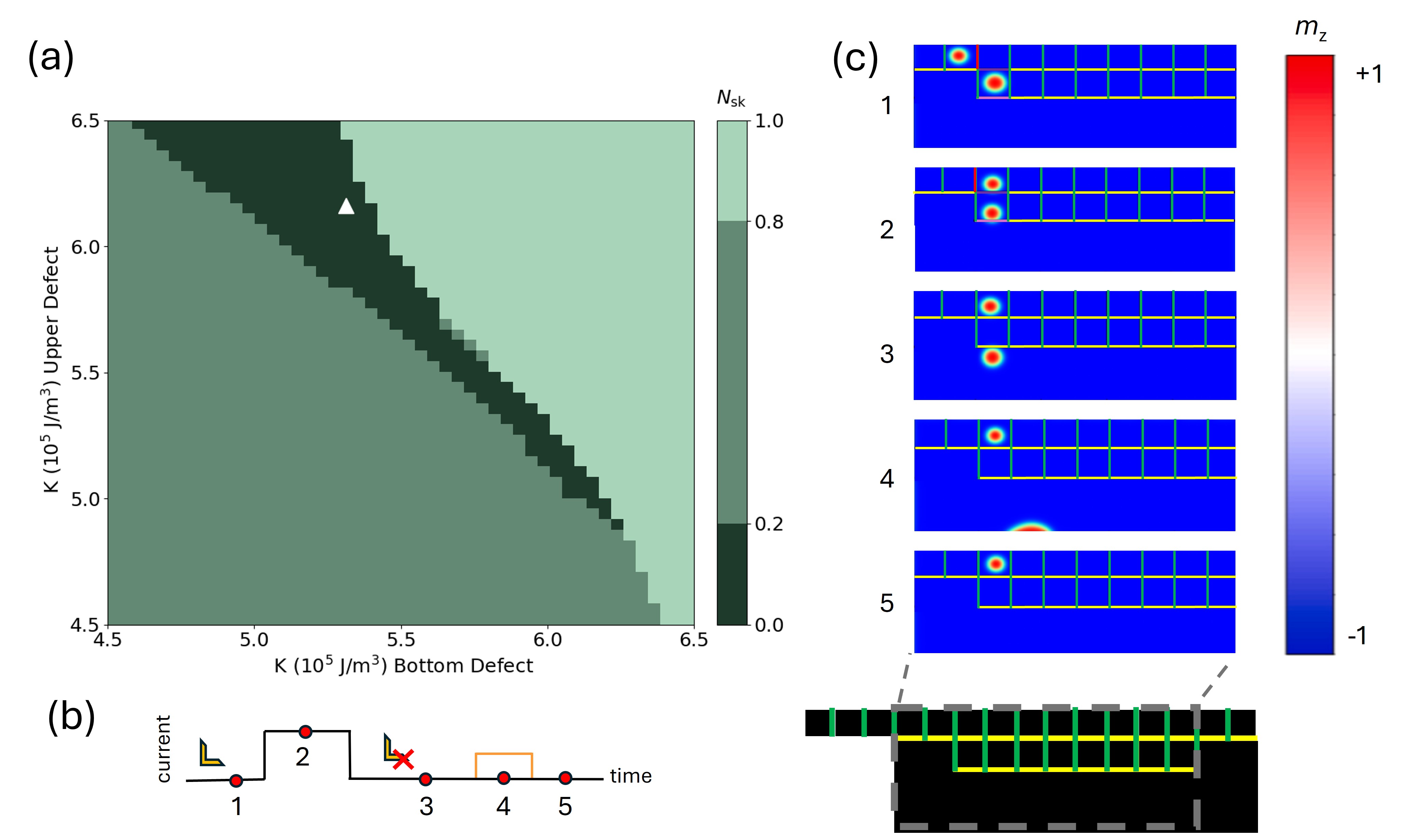}
      \caption{(a) Phase diagram for the skyrmionic number in a memory cell as a function of the values of the anisotropy of the upper and bottom borders of the $0$-th memory cell; the left border of the $0$-th carry cell is also modified at $1.15K_0$; initially, there is a skyrmion in the memory cell. (b) Sketch of the temporal protocol applied; the symbology is the same as the sketch in Fig.\ref{fig:1skyrmion}; the orange line indicates the application of a current $J_b$ in the buffer region, during a time $\Delta_b$ ($J_b=10J_H$ and $\Delta t_b=2.5 \mathrm{~ns}$ in this example)  (c) Snapshots of the $z$-component of the normalized magnetization at the instants indicated as red points in (b) using the values for anisotropies indicated with a triangle in (a).}
    \label{fig:barrier2}
\end{figure*}


Figures \ref{fig:barrier1} and \ref{fig:barrier2} show the phase diagrams for the skyrmionic number in the memory cell for several values of the modified anisotropy. The figures are interpreted as follows.

In Fig. \ref{fig:barrier1}, there is, initially, a skyrmion in the carry track and none on the memory track. We observe two main regions: the dark region indicates that the skyrmion in the carry track does not enter the memory track ---or enters and leaves---, while the light region indicates that the skyrmion has entered  the memory track and remains there. The snapshots in \ref{fig:barrier1}(c) correspond to this latter case. Note that the protocol used for the manipulation involves, not only current application, but also anisotropy modification.

In Fig. \ref{fig:barrier2} there is, initially, in addition to the skyrmion on the carry track, another one in the memory track. We observe three main regions: the darker region indicates that the skyrmion in the carry track has stayed there, but the skyrmion that was in the memory track has been expelled [the snapshots for this case are shown in Fig. \ref{fig:barrier2}(c)]. The lighter region corresponds to the case where the skyrmion in the carry track moves to the memory track, while the skyrmion in the memory track is expelled to the buffer region. Finally, the medium-light region corresponds to the case in which the skyrmion in the carry track has pushed the one on the memory track but, instead of occupying the memory track has been pinned in the intermediate region between the memory and the buffer track.

At this point, we would like to emphasize that opening or closing the "gates" according to the skyrmion position does not constitute a fully in-memory computing procedure. The gates are activated or deactivated based on the location of the skyrmion, meaning that the control operation relies on knowledge of the memory state. In a genuine in-memory computing, however, the operations should be completely blind to the positions of the memory skyrmions throughout the computation process.

\subsection{Annihilation}
\label{sec:anihilation}
Finally, the last primitive tool we are going to use is the annihilation of skyrmions. There are many ways in which this can be done. We do not focus on this step here. For the sake of simplicity, we just use a buffer region in which, when desired, all the skyrmions that are there are annihilated. There are many ways of doing that. As an example, one can apply a large enough current $J_b$ during a time $\Delta t_b$ in the buffer region in such a way that, when the skyrmions are there, they are forced to move towards the border of the buffer region and eventually be destroyed there. This is sketched with a red line in the protocol of Fig.~\ref{fig:barrier2}(b) using values $J_b=10J_H$ and $\Delta t_b=2.5 \mathrm{~ns}$. 

\section{Operation Protocols}
\label{sec:operations}

The skyrmions can be created, moved from cell to cell, and eventually annihilated, thus changing the stored value in the byte track. If the skyrmions are considered bits of information, the goal is to have protocols for different mathematical operations, independent of the actual information of the byte. That is, we would like to perform a blind operation where, without the need to read the input data, the output yields the desired operation. As a trivial example, consider a given value in the byte-track [i.e., $n = (00001011)_2 = (11)_{10}$, as in Fig.~\ref{fig:sketch}]. A doubling of this value is to move all the "1" in the binary number one position to the left ---all the skyrmions one cell to the right in the memory track [in Fig.~\ref{fig:sketch} this is $2n=(00010110)_2=(22)_{10}$]. So, the protocol for the blind operation "multiplication by 2" is "to move all skyrmions one cell to the right". 

Building on the primitive tools introduced in the previous section, we are now ready to present protocols for completely blind in-memory computation. We now define the problem precisely: consider a memory track whose bits encode a binary number $n$, where a bit value of 1 corresponds to the presence of a skyrmion and a bit value of 0 to its absence. We would like to have a physical protocol for the operation $f$ such that, using the defined primitive manipulation tools, without knowing where the 1's and 0's are, and without reading them at any moment during the computation, the final memory track bits encode the value $f(n)$.

We will define the protocols for two of the basic arithmetic operations: doubling: $f(n)=2n$, and successor: $f(n)=n+1$. Importantly, $n$ is unknown.

\subsection{Doubling protocol}
Doubling is the easiest operation with binary numbers. The protocol, in our binary memory-track would be "move all 1's and 0's one cell to the right". 
In fact, this is what we have already seen in section \ref{sec:manipulation} and in Fig.~\ref{fig:1skyrmion}. So, the "$\times 2$"-protocol, for the function $f(n)=2n$, would be simply: 
\begin{itemize}
\item[i.] Set the left-right borders anisotropy to an adequate value (Fig.~\ref{fig:1skyrmion}).
\item[ii.] Set a current $J_H$ in the memory track during a time $\Delta t_H$.
\end{itemize}
We have checked that the presence of neighboring skyrmions does not distort the behavior shown in Fig.~\ref{fig:1skyrmion} for a single skyrmion, within the whole range of parameters considered. 

Applying $m$ times this protocol we obtain the protocol for the function $f(n)=2^m \, n$. We note that, as the movement of skyrmions runs, small perturbations in the position could trigger a mistake in the operation, especially for large $m$'s, thus reducing the reliability of the operation. In this sense a safer (but slower) way of performing a $\times 2^m$ operation would be to wait a given time between successive applications of the $\times$2-protocol, for instance. Optimization of the protocol in terms of reliability and speed of operation is possible, but beyond the scope of this work.

It is important to note here that in the case of doubling or multiplication by 2$^m$, neither the carry track nor the buffer region are needed. Simply moving the skyrmions $m$ cells is enough to perform this blind operation. 

Also note that $m$ could be a negative number in the $f(n)=2^m\, n$ operation. In this case the protocol would be exactly the same but setting current in the opposite direction. In this case, if the number is odd (a 1 in the least significant bit-cell), we would need to have an extra bit-cell to hold the  $2^{-1}=0.5$ value. 

\subsection{Successor protocol}

This is one of the most fundamental operations. In our bit-cell platform, incrementing an unknown binary number by one can be described as follows: starting from the least significant bit, clear all consecutive 1's from the least significant bit up to the first $0$, and then place a $1$ in that cell.

We should elaborate a blind protocol to perform this operation. To do so, we take advantage of the repulsive interaction between skyrmions and the anisotropy manipulation described in the previous section. The idea is the following: after generating a skyrmion in the carry-track, we move it along the carry-track in such a way that when coinciding with a skyrmion in the memory-track, the repulsion between them would be enough so that the memory skyrmion passes the defect barrier and enters the buffer region, but not enough to move the carry-skyrmion from the carry-track [Fig. \ref{fig:barrier2}(a), dark region]. If there is no skyrmion in the memory-track, the carry-skyrmion will move to the memory-track and will stay there [Fig. \ref{fig:barrier1}(a), light region].  In this way, we have made the $f(n)=n+1$ operation. This procedure can be performed using variable barriers between cells according to the results of Figs. \ref{fig:barrier1} and \ref{fig:barrier2}. Again, we should note that the changes should not depend on the actual presence/absence of skyrmions but must be blind.

Summarizing, the physical protocol is the following:
\begin{enumerate}
    \item[i.] Generate a skyrmion in the carry region and move it  (section \ref{sec:manipulation}-C) to the $-1$ position of the carry track. Set $i=0$.
    \item[ii.] Modify the anisotropies of the left border of the $i$-th cell of carry track and of the upper and bottom borders of $i$-th memory cell.
    \item[iii.] Apply a current $J_H$ in the carry track during a time $\Delta t_H$. Wait a time $\Delta t_w$.
    \item[iv.] Reset the anisotropy of the modified barriers to their basis value. Annihilate the skyrmions of the buffer region (for example applying currents in the buffer region, section \ref{sec:manipulation}-D).
    \item[v.] Repeat the process from the point (ii.) increasing $i$ by 1, up to $i=7$ (no more bit-cells).
\end{enumerate}

\begin{figure*}
    \centering
    \includegraphics[width=0.8\linewidth]{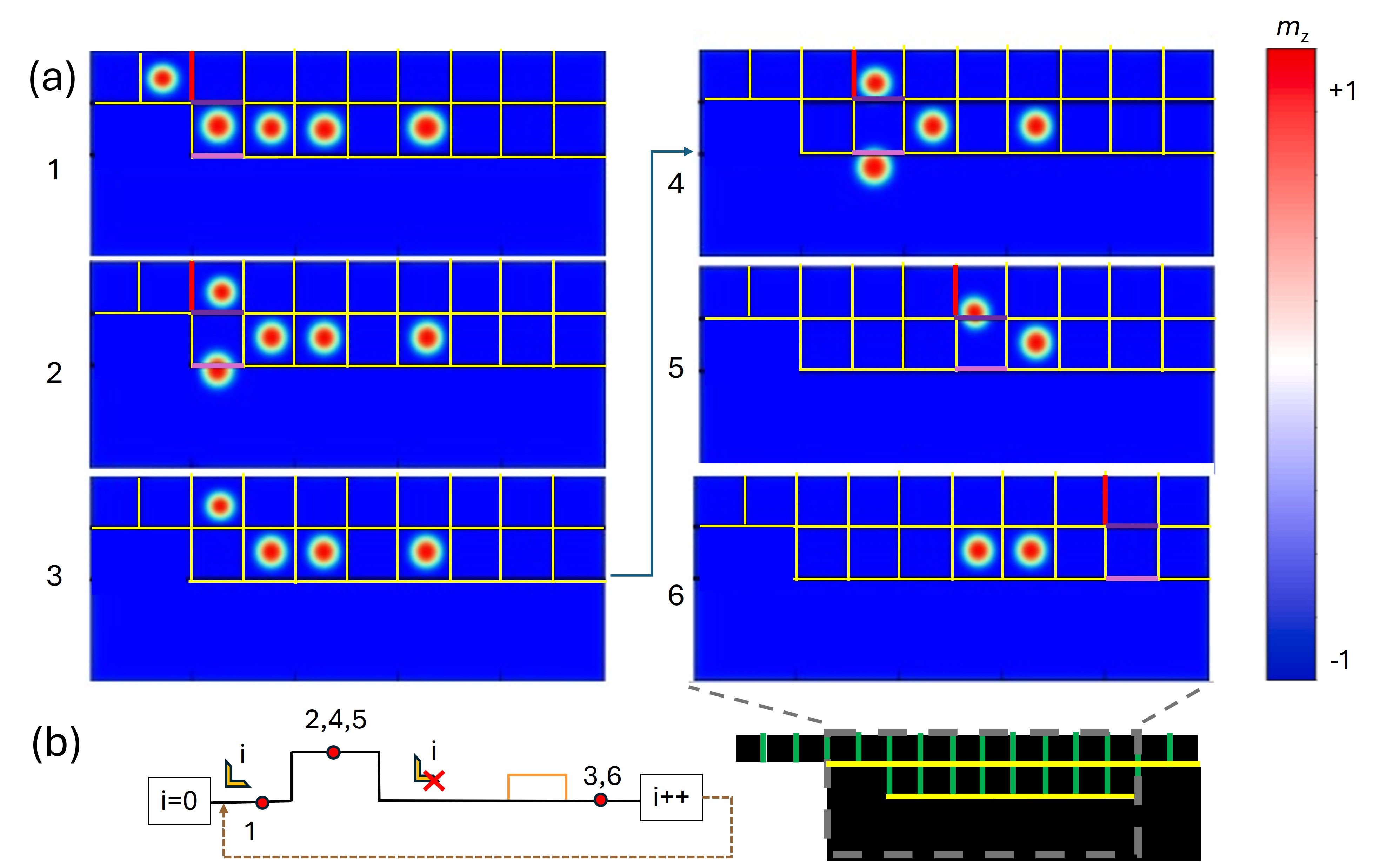}
    \caption{(a) Snapshots for the $z$-component of the magnetization during the successor $f(n)=n+1$ protocol. $J_H=8.19\times 10^{10} \mathrm{~A/m^2}$, $J_b=10J_H$, $\Delta t_H = \Delta t_w = \Delta t_b= 2.5\,\mathrm{ns}$, and the values of modified anisotropies at borders are those indicated by the triangles in Figs.\ref{fig:barrier1} and \ref{fig:barrier2}. Each snapshot corresponds to the moment indicated in (b) as red circles. (b) Sketch of the protocol followed for the successor function. Symbols are the same as in previous protocol figures; the loop (line with $i++$) indicates repetition of the process at the next cell. The initial and final snapshots correspond to the binary number $n=(00010111)_2=(23)_{10}$ and $n+1=(00011000)_2=(24)_{10}$, respectively.}
    \label{fig:plusone}
\end{figure*}

This protocol has been implemented in our system. Snapshots of the movement of the skyrmions during the application of the protocol are presented in Fig. \ref{fig:plusone} for the case of the number $(00010111)_2=(23)_{10}$. The values of the modified anisotropies at borders are selected following Figs.\ref{fig:barrier1}-a and \ref{fig:barrier2}-a (marked with triangles). A waiting time $\Delta t_w$ and a current in the buffer layer to annihilate the skyrmions present there are introduced to ensure that the system stabilizes before the next step begins (in the presented case, $\Delta t_w =\Delta t_b = 2.5 \,\mathrm{ns}$). Additionally, we have checked the protocol for several initial configurations in the memory track, but with exactly the same values of currents and anisotropies. In all cases, the final state corresponds to a value that is that of the initial state incremented by one, as intended. In the example, $(00011000)_2=(24)_{10}$.

It is worth noting that once the carry skyrmion reaches the memory track, no further changes occur. However, since the computation must be fully blind and the entry point of the skyrmion into the memory track is not known \emph{a priori}, the protocol must be executed in its entirety to guarantee correct operation.

This procedure depends on the values of $J_H$, $\Delta t_H$, $\Delta t_w$, and the barrier modification factor. All these values can vary in a range of values and depend on the cell size and the magnetic parameters of the system. Further optimization of the geometry, barrier profiles, current amplitudes, and timing parameters could enhance both the robustness and the practical feasibility of the system.



\section{Discussion and conclusions}
\label{sec:discussion}
We have presented several protocols for genuine in-memory computing using binary numbers physically encoded by the presence (1) or absence (0) of skyrmions. Computation is performed by directly manipulating the skyrmions that act as memory bits. In this way, operations can be carried out blindly, without requiring knowledge at any stage of the computation of the actual positions of the skyrmions. This approach avoids energetically costly read/write procedures during intermediate steps, requiring only initial writing and final readout.

Skyrmions, being excellent candidates for spintronic magnetic memories, also offer strong potential for direct in-memory computing. In this work, we demonstrate this potential through simulations showing how such an in-memory skyrmionic platform could be implemented. Despite this theoretical potential, several challenges must be addressed in order to move from proof-of-concept to practical implementation:
\begin{itemize}
\item[(i)] \textit{Temperature:} In our simulations, the temperature has been set to zero. However, it is well known that finite temperature can destabilize skyrmions and induce diffusive motion. For practical implementations, materials in which skyrmions are stable at room temperature should be employed \cite{Matsumoto2018-NanoLett}. Additionally, temperature effects are expected to reduce the range of valid parameters due to a weakening of defect barriers. Thermal fluctuations may also induce random motion of skyrmions \cite{Wang2022-NatComm}. In our platform, however, skyrmions are confined by engineered barriers and defects, which should significantly mitigate these effects.
\item[(ii)] \textit{Defects:} Material imperfections must also be taken into account \cite{Del-Valle2022-APLMat}. In our model, we assume that the influence of intrinsic defects and grain boundaries is much smaller than that of intentionally introduced and controlled defect barriers. Random defect-induced pinning is neglected under the assumption that the applied driving currents are sufficiently large to overcome such effects.
\item[(iii)] \textit{Optimization:} The system considered here is designed to demonstrate the feasibility of skyrmion-based in-memory computing. The chosen parameter ranges are not pushed to extreme limits. Further optimization is possible, both in terms of device geometry and physical parameters, and is expected to significantly improve performance.
\end{itemize}

To meaningfully compare the proposed platform with existing computational technologies, one would need to evaluate metrics such as computation time, energy consumption, feasibility, error tolerance, cyclability, and others. This is much beyond the scope and purpose of the present work. Rather than proposing a competitive computing architecture, our goal is to introduce a novel platform for in-memory computing with the potential to drastically reduce energy consumption, once optimized.
This perspective is supported by the intrinsic properties of skyrmions: they are nanoscale objects, can be moved at high speeds, and require low energy for manipulation. Consequently, a fully in-memory skyrmionic platform could offer high storage density, fast operation, and low power consumption.

\begin{acknowledgments} 
We thank Samuel Del Valle for discussions. We also acknowledge financial support from the Spanish Ministry of Science, Innovation and Universities MCIU/AEI through PID2023-149054NB-I00. J.C.-S.~acknowledges funding from AGAUR-FI Joan Or\'o grants (Grant No. 2025 FI-3 00143) of the Generalitat de Catalunya. 
\end{acknowledgments}

\bibliography{bib_inmemorysky} 

\end{document}